\documentclass[16pt,a4]{article}
\usepackage{epsfig}
\usepackage{setspace}
\title{Reaction Time of a Group of Physics Students}
\author{Charu Saxena, Rini Kaur \& P.Arun\\
Department of Physics and Electronics,\\
S.G.T.B. Khalsa College\\
University of Delhi, Delhi 110 007, India.}

\begin{document}
\maketitle
\begin{abstract}
The reaction time of a group of students majoring in Physics is reported here. 
Strong co-relation between fatigue, reaction time and performance have been 
seen and may be useful for academicians and administrators responsible of 
working out time-tables, course structures, students counsellings etc.
\end{abstract}

\section*{Introduction}
Animal's respond to the environment using their sensory organs for 
collecting information that is passed on to the brain and analyzed for action. 
However, this would take a perceivable time. This time is called the reaction 
time. The definition of reaction time or latency as given in the 
wikipedia is ``{\sl the time from the onset of a stimulus until the 
organism responds}'' \cite{r2}. Human reaction time is ultimately limited by 
how fast 
nerve cells conduct nerve impulses. Although this speed is almost 250 miles 
per hour, messages still take a significant amount of time to travel from 
sensory organs to the brain and back to the appropriate muscle groups.

A common ``experiment'' done as a game by children is for one boy to hold a 
scale about chest high and have someone place his thumb and index finger 
about an inch apart somewhere along the (bottom) length of the scale. Now, he 
would have to catch the scale when the first boy allows it to fall. The scale 
won't be able be caught immediately and a length of the scale would pass 
through his finger before it is caught. From simple laws of mechanics, using 
the equation
\begin{eqnarray}
t=\sqrt{{2 s \over g}}\nonumber
\end{eqnarray} 
the reaction time of the child can be calculated.
\begin{figure}[t!!]
\begin{center}
\includegraphics[width=3.5in,angle=-90]{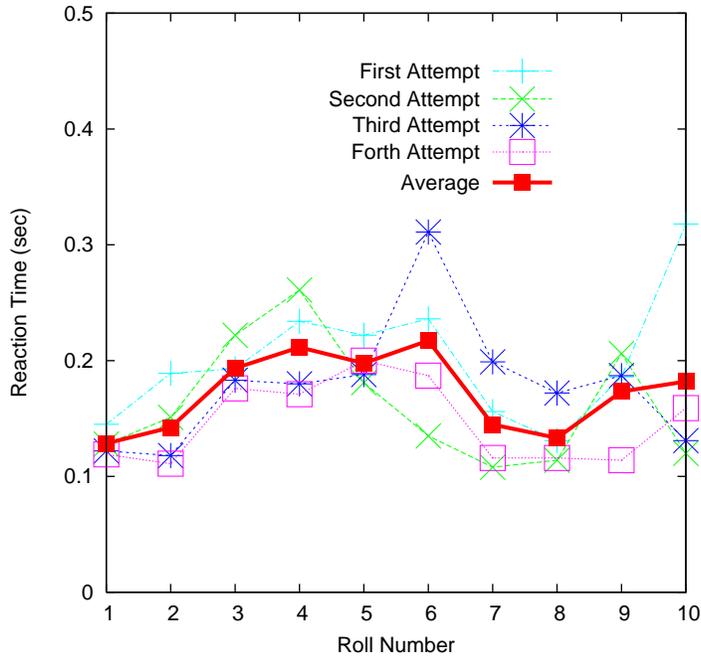}
\caption{The dark thick line shows the variation of the average time recorded 
by the students of the given roll number. The repeated attempts of the 
students resulted in an average ${\pm 0.422s}$ deviation from the their 
average.}
\end{center}
\end{figure}
Interest in the measurement of human reaction time 
apparently began as a result of the work of a Dutch physiologist named 
F.C.Donders. Beginning in 1865, Donders \cite{d1} became interested in the 
question of 
whether the time taken to perform basic mental processes could be measured. 
Until that time, mental processes had been thought to be too fast to be 
measurable. In his early experiments, Donders applied electric shocks to the 
right and left feet of his subjects. The subject's task was to respond by 
pressing a telegraph key with his right or left hand to indicate whether his 
right or left foot had received the shock.
Interest in measuring and minimizing the reaction time today is of interest in 
medicine, military, traffic control and sports. Things can be put to better 
perspective by taking an example. In the game of 
cricket, the average distance between the bowler and batsman is 20mtrs. With a 
spin bowler delivering the ball at around 80Km/hr, the batsman has 0.9s (900ms)
to ``see'' the ball, decide the shot and implement it! An analysis of 
high-speed 
film of international cricketers batting on a specially prepared pitch which 
produced unpredictable movement of the ball is reported, and it is shown that, 
when batting, highly skilled professional cricketers show reaction times of 
around 200ms \cite{r1}.

Various methods have been used to measure the 
reaction time. Essentially, measuring simple reaction time like in Donders 
experiment or recognition reaction time or choice reaction time. 
In choice reaction time experiments the subject must give a response that 
corresponds to the stimulus, such as pressing a key corresponding to a letter 
as soon as the letter appears on a display amist random display of characters. 
In this article we are reporting 
the results of our experiment done using this method. 
The reaction time is known to be effected by factors such as age, gender, 
fatigue/ exercise, distractions and intelligence. Our sample group were 
students of physics/ electronics in the age group of 18-21, where studies 
have shown the reaction time to be the minimum in a human life span 
\cite{galton,f1,wel1, wel2,bel1}. These works report the reaction time of 
people in the age group of our study to be $\sim$200ms. 

\section*{Results}
\begin{figure}[b!!]
\begin{center}
\includegraphics[width=2in,angle=-90]{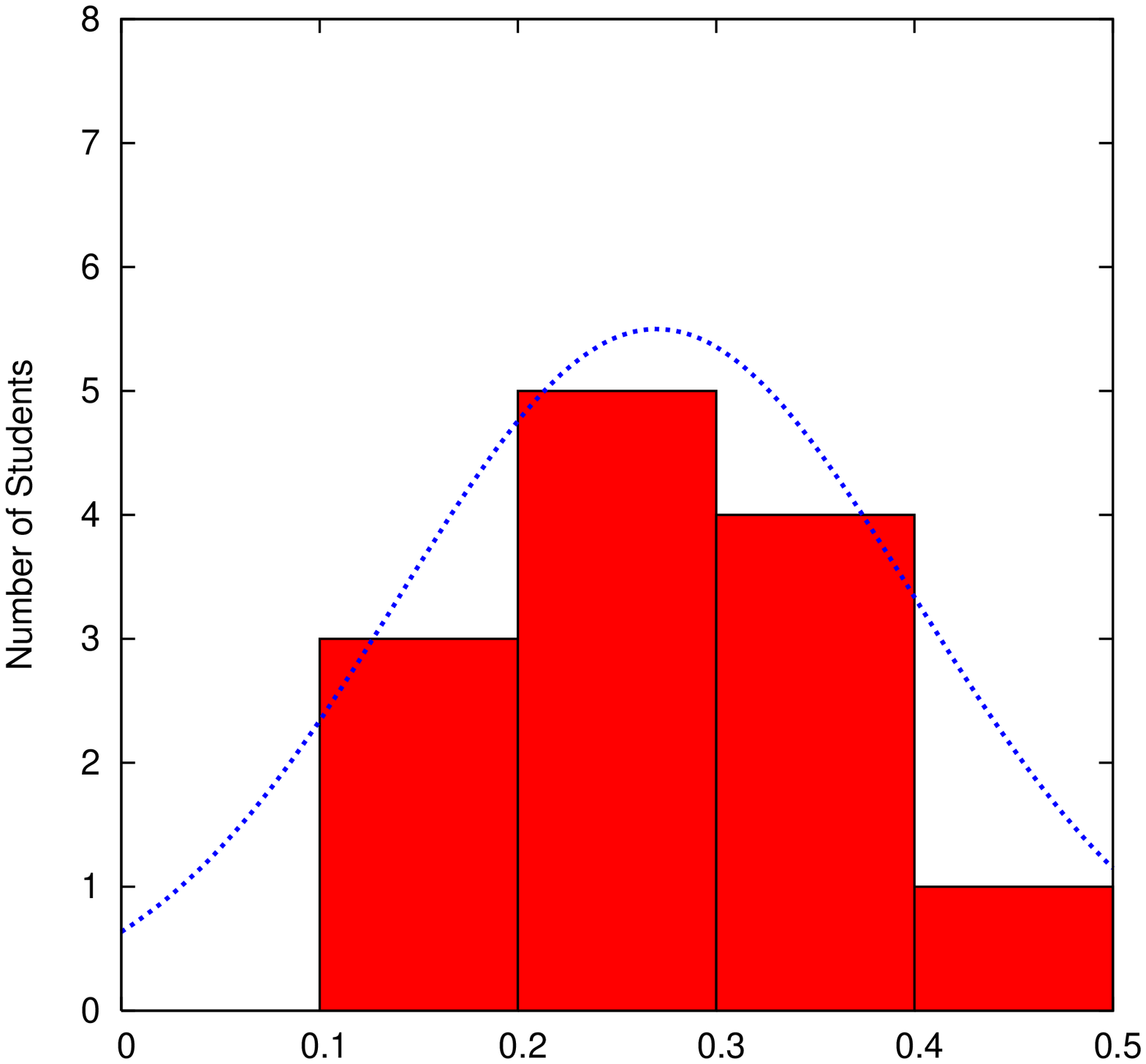}
\includegraphics[width=2in,angle=-90]{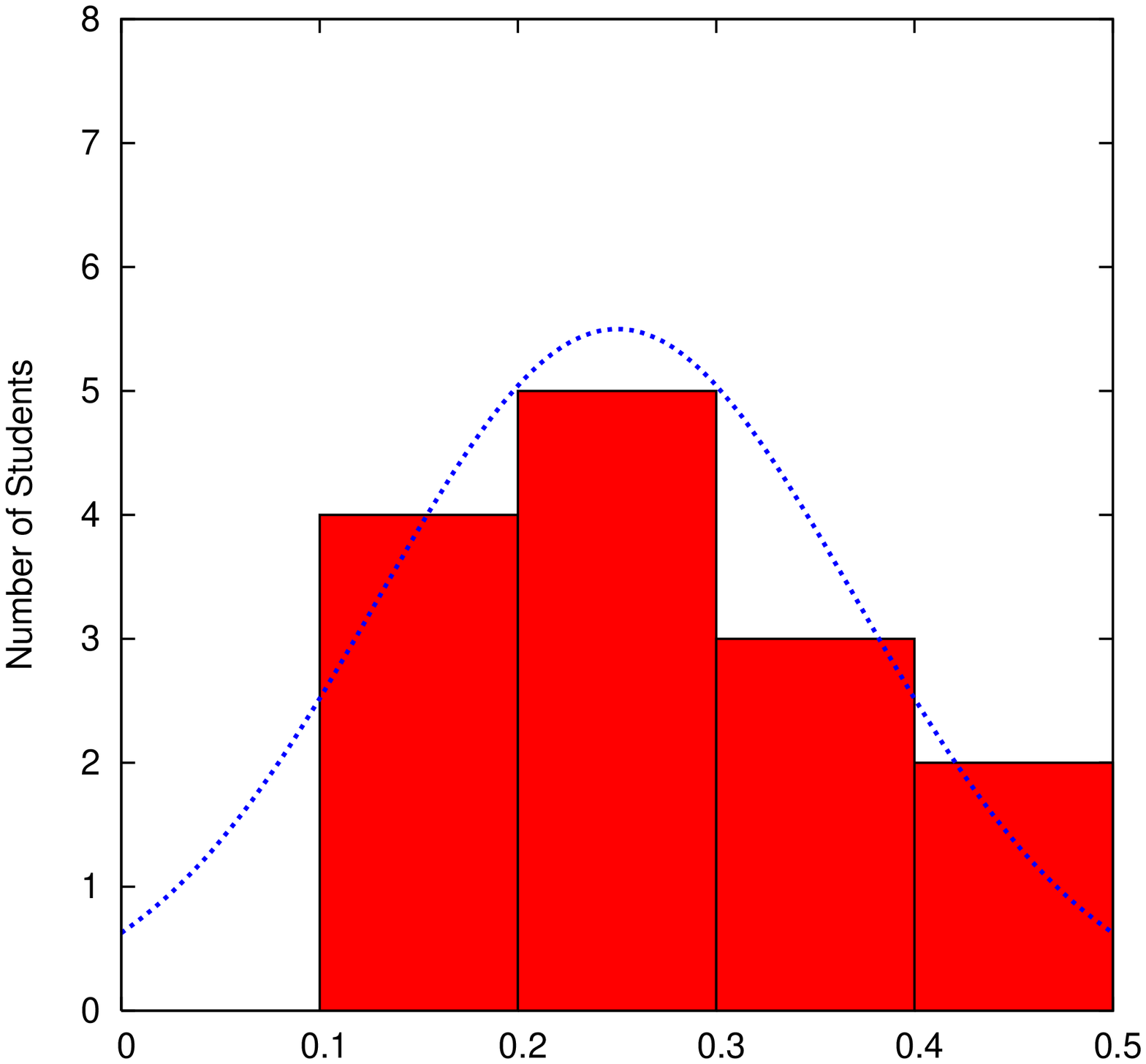}
\includegraphics[width=2in,angle=-90]{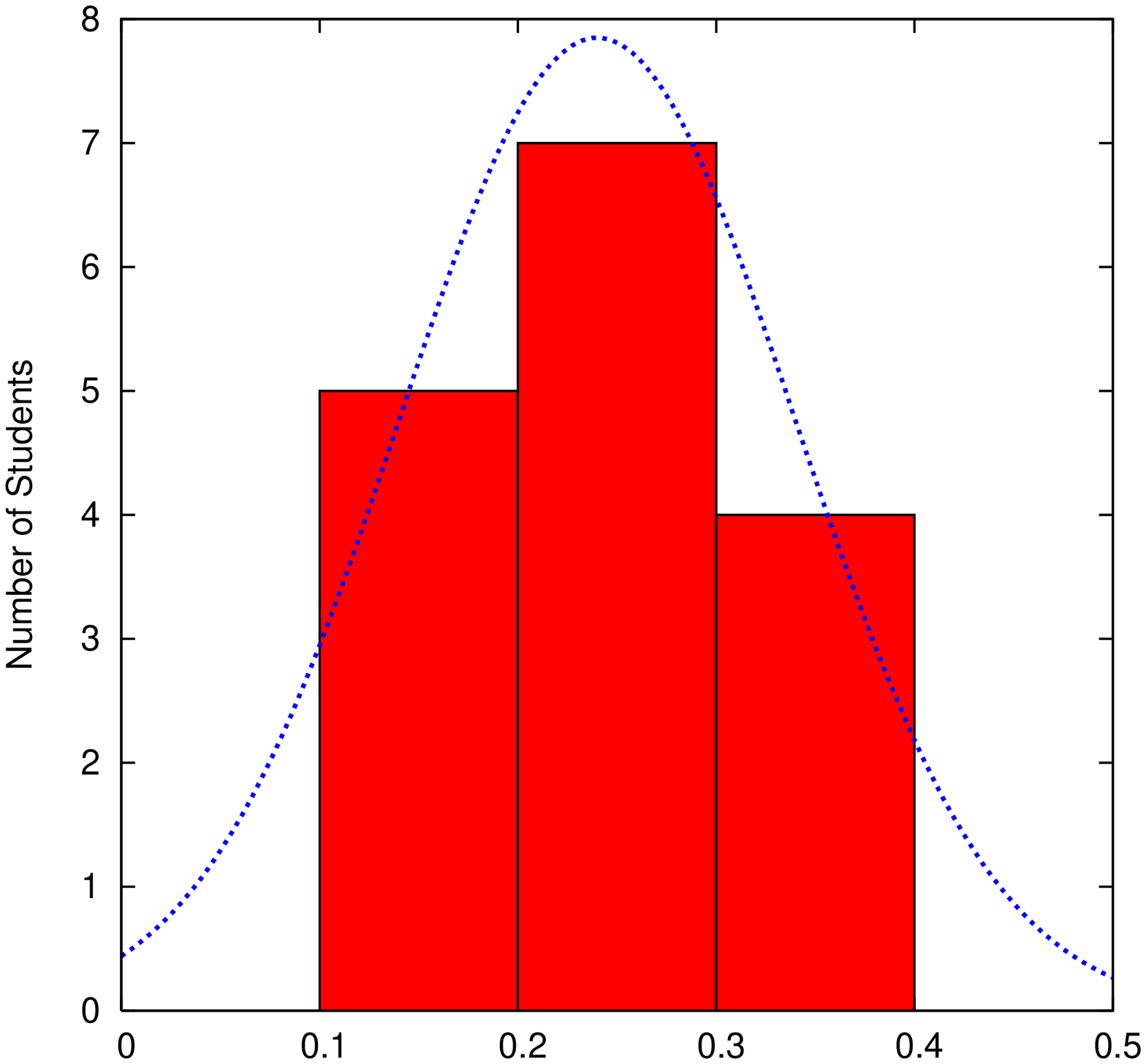}
\caption{The distribution of the reaction time of the ${\rm 1^{st}year}$, 
${\rm 2^{nd}year}$ and ${\rm 3^{rd}year}$ majoring in physics/ electronics.}
\end{center}
\end{figure}
We have sampled and recorded the reaction time of 137 students, however, here 
we discuss data of 44 students who were majoring in physics/ electronics. 
Usually such experiments sample 20 or more people and make them  
repeat the experiment over large time \cite{Luce}. Another approach is 
where single reading is taken after allowing the test person a period of 
practice \cite{Sanders}. We first tested the effect of practice on a group of 
students. Fig~1 shows a the variation of reaction time of students with 
increasing practice. Raw reaction time, i.e. the first attempt, of the 
students were poor. As they practiced, a recording was taken at every 
15 minutes. 
Practice however did not keep on improving the reaction time. Only four out of 
ten students had better reaction time on their forth recording (i.e. after 45 
minutes of practice) as compared to 
their third try (30 minutes into practice). While one might be tempted to 
conclude this as improvement 
with practice, it should be noted that three of these four students in their 
forth attempt performed worse then their second attempt. The spectrum of 
reaction time is within ${\rm \pm 0.422s}$ of the average values. This 
deviation is just ${\rm \pm 0.075s}$ when the first attempt is neglected.
Hence, in our experiment, our approach has been to allow a subject to 
familiarize the machine for 20-25 minutes before taking their reaction time.
\begin{table}[h]
\begin{center}
\caption{Table compares the male-female distribution of the three classes and 
their preformances.}
\vskip 0.5cm
\begin{tabular}{|c|c| c| c| c| }
\hline
Class & Girls & Boys & c & Marks Obtained \\ \hline
${\rm 1^{st}y}$ & 6 & 8 & 0.13 & 69 \\ \hline
${\rm 2^{nd}y}$ & 4 & 14 & 0.12 & 74  \\ \hline
${\rm 3^{rd}y}$ & 9 & 5 & 0.10 & 87  \\ \hline
\end{tabular}
\label {tab}
\end{center}
\end{table}

Fig~2 shows the performance of the students from the first, second and third 
years majoring in Physics/ Electronics. Along with each histogram, a Gaussian 
\begin{eqnarray}
f(x)=ae^{-{(x-b)^2 \over 2c^2}}
\end{eqnarray}
was fitted to estimate the mean reaction time (b) of the class and the 
deviation from the mean (using c). Table~1 details the results for all three 
classes. The lower mean reaction time and narrower deviation from the mean 
of the third year students show a collective better performance. A boarder 
sampling of reaction time with larger age variation was collected based on 
gender (results not shown here). We found no variation in performance based 
on gender with the ratio of 
female to male reaction time being equal to unity, i.e. (${\rm R_{FM}=1}$). 
Bellis \cite{bell1} and Engel \cite{angel} reported ${\rm R_{FM}=1.1}$, with 
males having a faster reaction time. However, recent studies by Silverman 
\cite{silver} reports the difference in male-female reaction time was 
narrowing. Table~1 gives the number boys and girls in each class. Eventhough 
ratio of boys and girls are not same in these classes, no correction is called 
for in fig~3 since ${\rm R_{FM}=1}$. 

\begin{figure}[h!!]
\begin{center}
\includegraphics[width=3.5in,angle=-90]{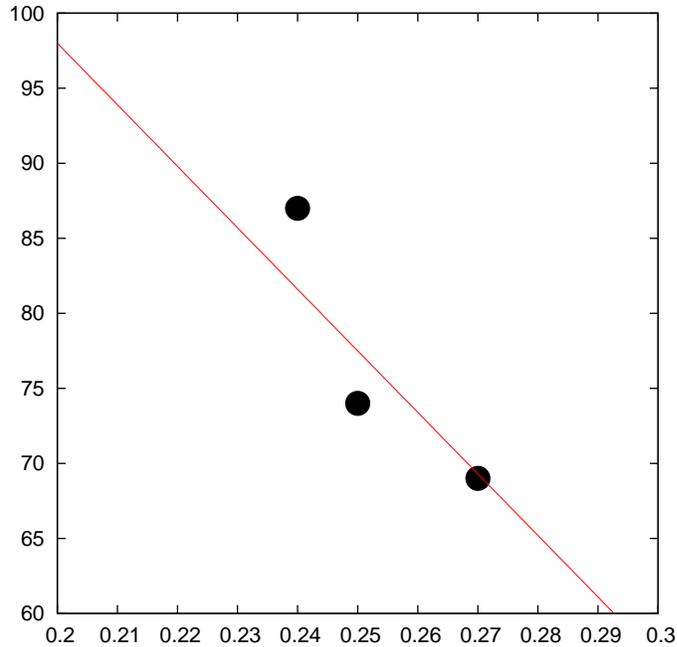}
\caption{Comparing the performance of the students of ${\rm 1^{st}}$, 
${\rm 2^{nd}}$ and ${\rm 3^{rd}}$ year in an exam as a function of their 
class's mean reaction time. The line is only to show trend.}
\end{center}
\end{figure}

As stated earlier, the ${\rm 3^{rd}year}$ students appear to be a sharper lot 
and it was thought worthwhile to test if the reaction time had any co-relation 
with learning ability. A comprehensive test was designed to test all the 
students under study for their ability to comprehend, learn on their own, 
analyze and solve a given problem. The test was different from the ordinary 
annual examination these students face and also care was taken that the 
evaluator's are not prejudiced or influenced by the results of fig~2. 
Fig~3 shows how reaction time seem to co-relate strongly with the student's 
ability to learn. This result is consistent with the findings of Deary et 
al \cite{dear}. All these students were admitted to the college based on 
their performance in higher secondary (HS) examination conductedd by CBSE 
(India). All the students had marks between 78-84\% in their HS examination. 
The resolving of their performance with respect to their reaction time hence 
was made possible because of the complex method adopted for evaluation. 
Schweitzer \cite{sch} in his paper reports that the speed advantage of more 
intelligent 
people is greatest in tests requiring complex responses. In corollary, fig~3 
suggests sharper (faster) students are stimulated and respond keenly to tests 
having a degree of complexity.

\begin{figure}[t!!]
\begin{center}
\includegraphics[width=4.5in,angle=0]{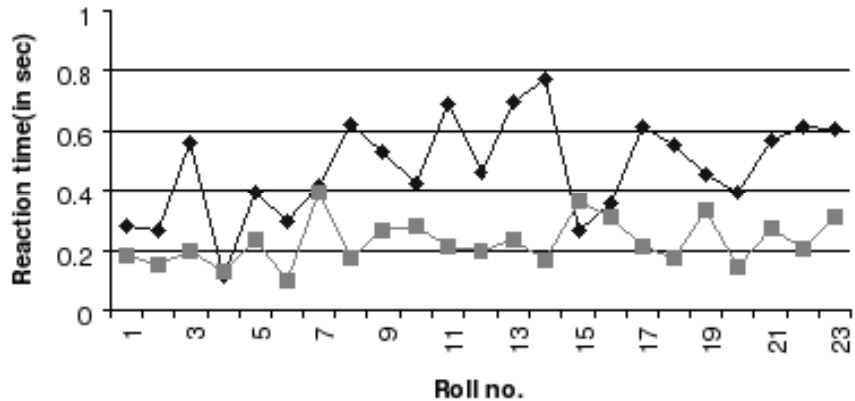}
\caption{The variation in reaction time after the students went through a 
``stressful'' hour of theory class in Mathematical Physics. Clearly their 
reflexes were found wanting after the class.}
\end{center}
\end{figure}

Another interesting result we have is how the students reaction time varies 
after they attend an hour of tendious lecture. We took recordings of the 
first year students before they entered the lecture hall and again, an hour 
later as they emerged from their Mathematical Physics class. The particular 
subject was selected after studying their response to a questionaire, where 
majority reported it as the most difficult subject. The subject was also low 
on popularity. Also, their performance in class tests were consistently poor. 
Fig~4 shows the variation of performance. Barring four students, all the 
students showed deterioration in reaction. {\sl ``Stress''}, hence, makes 
the reaction time poor. Slower response due to fatigue doing complicated 
task was reported as back as in 1953 \cite{single}.

\section*{Conclusion}
In conclusion, the fatigue level seen in students after attending an hour of 
intense 
training in Mathematical Physics does advocate reducing the duration of a 
lecture from 60 minutes to 40-45 minutes for students below 21 years. A 
reduction in response time is also a reduction in concentration level and hence
suggests much of the information imparted by the instructor (Fig~4) would 
have anyway not been absorbed. The methodolgy adopted in teaching difficult 
subjects also should be reviewed\footnote{This statement is reserved for 
methodolgy followede in Delhi University, since the authors are in no position 
to comment on senarios elsewhere.}.

The Gender myth that girls are poor in subjects like maths and engineering 
subjects were also broken in this sample of study.


\begin{thebibliography}{}
%
\bibitem{r2} http://biae.clemson.edu/bpc/bp/Lab/110/reaction.htm

\bibitem{d1} F. C. Donders,{\sl ``On the speed of mental processes,''} 
Translated by W. G. Koster, {\bf Acta Psychologica 30} (1969) 412-431.

\bibitem{r1} P. McLeod, {\sl ``Visual reaction time and high-speed ball 
games,}'', {\bf Perception, 16} (1987) 49-59.

\bibitem{galton} F. Galton, ``{\sl On instruments for (1) testing 
perception of differences of tint and for (2) determining reaction time.}'', 
{\bf Journal of the Anthropological Institute, 19} (1899) 27-29.

\bibitem{f1} K. von Fieandt, A. Huhtala, P. Kullberg, and K. Saarl,
``{\sl Personal tempo and phenomenal time at different age levels}'', 
{\bf Reports from the Psychological Institute, 2} (1956).

\bibitem{wel1} A. T. Welford, ``{\sl Motor performance. In J. E. Birren and 
K. W. Schaie (Eds.), Handbook of the Psychology of Aging}'' 
Van Nostrand Reinhold, New York (1977) pp. 450-496.

\bibitem{wel2} A. T. Welford, ``{\sl Choice reaction time: Basic concepts.}'' 
In A. T. Welford (Ed.), Reaction Times. Academic Press, New York (1980), 
pp. 73-128.

\bibitem{bel1} J. T. Brebner, and A. T. Welford, ``{\sl Introduction: an 
historical background sketch}'', In A. T. Welford (Ed.), Reaction Times. 
Academic Press, New York (1980), pp. 1-23.

\bibitem{Luce} R. D. Luce, ``{\sl Response Times: Their Role in Inferring 
Elementary Mental Organization}'', Oxford University Press, New York (1986).

\bibitem{Sanders} A. F. Sanders, ``{\sl Elements of Human Performance: 
Reaction Processes and Attention in Human Skill}''. 
Lawrence Erlbaum Associates, Publishers, Mahwah, New Jersey (1988).

\bibitem{bell1} C. J. Bellis, ``{\sl Reaction time and chronological age.}'', 
{\bf Proceedings of the Society for Experimental Biology and Medicine 30} 
(1933) 801.

\bibitem{angel} B. T. Engel, P. R. Thorne, and R. E. Quilter, ``{\sl On the 
relationship among sex, age, response mode, cardiac cycle phase, breathing 
cycle phase, and simple reaction time.}'', {\bf Journal of Gerontology 27}, 
(1972) 456-460.

\bibitem{silver} I. W. Silverman, ``{\sl Sex differences in simple visual 
reaction time: a historical meta-analysis (sports events).}''. {\bf Sex 
Roles: A Journal of Research 54}, (2006) 57-69.

\bibitem{dear} I.J. Deary, G. Der and G. Ford, ``{\sl Reaction times and 
intelligence differences: A population-based cohort study.}'', {\bf 
Intelligence 29}, (2001) 389.

\bibitem{sch} K. Schweitzer, ``{\sl Preattentive processing and cognitive 
ability.}'', {\bf Intelligence 29} (2001) 169.

\bibitem{single} W. T. Singleton, ``{\sl Deterioration of performance on a 
short-term perceptual-motor task.}'', In W. F. Floyd and A. T. Welford (Eds.), 
Symposium on Fatigue. H. K. Lewis and Co., London (1953), 163-172.

\end{thebibliography}
\end{document}